\documentclass[1p,times]{elsarticle}
\usepackage[utf8]{inputenc}

\usepackage[colorlinks,
linkcolor=blue, 
citecolor=blue,
urlcolor=blue
]{hyperref}

\usepackage[
separate-uncertainty=false,
multi-part-units=single,
inter-unit-product=\cdot,
product-units=power
]{siunitx}

\usepackage{amsmath}
\usepackage{amssymb}
\usepackage{amsfonts}
\usepackage{isotope}
\usepackage{graphicx}
\usepackage{subfigure}
\usepackage{placeins}
\usepackage[T1]{fontenc}
\usepackage{fancyhdr}
\usepackage{parskip} 
\usepackage{makecell}
\usepackage{booktabs}
\usepackage{makecell,multirow,diagbox}
\journal{Nuclear Instruments and Methods in Physics Research Section A}
\begin{document}

\title{Beam test of a \SI{180}{nm} CMOS Pixel Sensor for the CEPC vertex detector}
\author[a,c,1]{Tianya Wu}
\author[a,b,1]{Shuqi Li}
\author[a,c]{Wei Wang}
\author[a,b]{Jia Zhou}
\author[a,b]{Ziyue Yan}
\author[e]{Yiming Hu}
\author[e]{Xiaoxu Zhang}
\author[a,c]{Zhijun Liang\corref{cor}}
\author[a,c]{Wei Wei\corref{cor}}
\author[a,c]{Ying Zhang\corref{cor}}
\author[d]{Xiaomin Wei}
\author[a,b]{Xinhui Huang}
\author[e]{Lei Zhang}
\author[e]{Ming Qi}
\author[a,b]{Hao Zeng}
\author[a,b]{Xuewei Jia}
\author[a,c]{Jun Hu}
\author[a,c]{Jinyu Fu}
\author[a,b,c]{Hongyu Zhang}
\author[a]{Gang Li}
\author[a]{Linghui Wu}
\author[a,b,c]{Mingyi Dong}
\author[a,c]{Xiaoting Li}
\author[g]{Raimon Casanova}
\author[f]{Liang Zhang}
\author[f]{Jianing Dong}
\author[d]{Jia Wang}
\author[d]{Ran Zheng}
\author[a,c]{Weiguo Lu}
\author[g,h]{Sebastian Grinstein}
\author[a]{Jo\~{a}o Guimar\~{a}es da Costa}

\fntext[t1]{These authors contributed equally to this article.}
\cortext[cor]{Corresponding author: \href{mailto:liangzj@ihep.ac.cn}{liangzj@ihep.ac.cn} (Zhijun Liang), \href{mailto:zhangying83@ihep.ac.cn}{zhangying83@ihep.ac.cn} (Ying Zhang), \href{mailto:weiw@ihep.ac.cn}{weiw@ihep.ac.cn} (Wei Wei) }
\address[a]{Institute of High Energy Physics, Chinese Academy of Sciences, 19B Yuquan Road, Shijingshan District, Beijing, 100049, China}
\address[b]{University of Chinese Academy of Sciences, 19A Yuquan Road, Shijingshan, Beijing 100049, China}
\address[c]{State Key Laboratory of Particle Detection and Electronics, Beijing, 100049, China}
\address[d]{Northwestern Polytechnical University, Xi'an, China}
\address[e]{Department of Physics, Nanjing University, Nanjing  210093, China}
\address[f]{Institute of Frontier and Interdisciplinary Science and Key Laboratory of Particle Physics and Particle Irradiation, Shandong University, Qingdao, China}
\address[g]{Institut  de F\'{i}sica d'Altes Energies (IFAE), Bellaterra (Barcelona), Spain }
\address[h]{Catalan Institution for Research and Advanced Studies (ICREA), Barcelona, Spain}
\begin{abstract}
The proposed Circular Electron Positron Collider (CEPC) imposes new challenges for the vertex detector in terms of pixel size and material budget. A Monolithic Active Pixel Sensor (MAPS) prototype called TaichuPix, based on a column drain readout architecture, has been developed to address the need for high spatial resolution. In order to evaluate the performance of the TaichuPix-3 chips, a beam test was carried out at DESY \uppercase\expandafter{\romannumeral2} TB21 in December 2022. Meanwhile, the Data Acquisition (DAQ) for a muti-plane configuration was tested during the beam test.
This work presents the characterization of the TaichuPix-3 chips with two different processes, including cluster size, spatial resolution, and detection efficiency. The analysis results indicate the spatial resolution better than \SI{5}{\um} and the detection efficiency exceeding \SI{99.5}{\%} for both TaichuPix-3 chips with the two different processes.
\end{abstract}
\maketitle
\section{Introduction}
The Circular Electron Positron Collider (CEPC) is a frontier high-energy research facility complex proposed by the Chinese particle physics community. Its primary objective is to precisely measure the physical properties of the Higgs, $W$, and $Z$ bosons, as well as to search for new physics beyond the standard model~\cite{CEPC-SPPC-accelerator,CEPC-SPPC-detector}. One of the key challenges faced by the CEPC is the development of the vertex detector, which must provide a high spatial resolution better than \SI{3}{\um} for the innermost layer while maintaining a low power consumption below $\qty{50}{\mW\per\square\cm}$ to enable forced air-cooling. Additionally, a low material budget below \SI{0.15}{\%} $X/X_0$is required for each detector layer to minimize the effects of multiple scattering. As an intermediate milestone, the current goal is to complete a baseline vertex detector prototype, which can achieve a spatial resolution better than \SI{5}{\um} and a detection efficiency of \SI{99}{\%}.

A prototype of the Monolithic Active Pixel Sensor (MAPS), named TaichuPix, has been developed for the CEPC baseline vertex detector~\cite{Wei:2019wbr,Wu:2021mju,Zhang:2022rlo}. It is based on a \SI{180}{\nm} CMOS Imaging Sensor (CIS) technology, the pixel pitch is \qtyproduct{25x25}{\um}, including a sensing diode region of \qtyproduct{8.6x8.6}{\um} and fast in-pixel readout electronics in the remaining area. Additionally, a dedicated high data rate peripheral readout logic has been implemented to support both trigger and triggerless modes. To meet the CEPC requirements for operating at the Higgs, $Z$, and $WW$ modes with branch-crossing intervals of \qty{680}{\ns}, \qty{25}{\ns}, and \qty{210}{\ns}, a maximum hit rate of \qtyproduct{36e6}{\per\square\cm\per\second} is required.
The power consumption of TaichuPix-3 is adjustable based on the event rate and running mode. When operating at a fast leading edge (< \qty{200}{\ns}) of the analog front-end and a serializer interface of \qty{160}{\MHz}, the power consumption is less than $\qty{200}{\mW\per\square\cm}$.

Three prototype versions have been designed: TaichuPix-1 and TaichuPix-2~\cite{Wu:2021mju,Zhang:2022rlo}, which are multi-project wafers, and TaichuPix-3, a full-scale prototype with an engineering run. The pixel matrix of TaichuPix-3 is \numproduct{1024x512} with a thickness of \qtyproduct{150}{\um}. The proposed baseline vertex detector consists of three layers of ladders~\cite{CEPC-SPPC-detector}, with double-sided mounted TaichuPix-3 sensors. In December 2022, a testing system consisting of six TaichuPix-3 sensors was set up in DESY \uppercase\expandafter{\romannumeral2} Test Beam Line 21 (TB~21)~\cite{DIENER:2019265}, which is one of three independent beamlines. Two different detectors under test (DUT) were evaluated in this test beam. DUT$_A$ is fabricated using the standard back-bias diode process, along with an extra deep N-layer mask~\cite{W:2017technogy}, DUT$_B$ is fabricated without the extra deep N-layer, serving as a comparison to DUT$_A$. It is noted that DUT$_A$ exhibits a larger depletion region than DUT$_B$ as mentioned in~\cite{W:2017technogy}.

 In this paper, the test beam results for DUT$_A$ and DUT$_B$ are discussed in terms of cluster size, spatial resolution, and detection efficiency. In addition, the DAQ for a muti-plane configuration is tested. The organization of this paper is as follows: the experimental setup for the beam test is described in Section~\ref{sec:ExperimentalSetup}, while the offline analysis and results are discussed in Section~\ref{sec:Offline analysis and results}. The summary and outlook are presented in Section~\ref{sec:conclusion}.

\FloatBarrier 
\section{Experimental Setup}\label{sec:ExperimentalSetup}
\subsection{General Setup}
\begin{figure}[ht]
    \centering
    \includegraphics[width=1\textwidth]{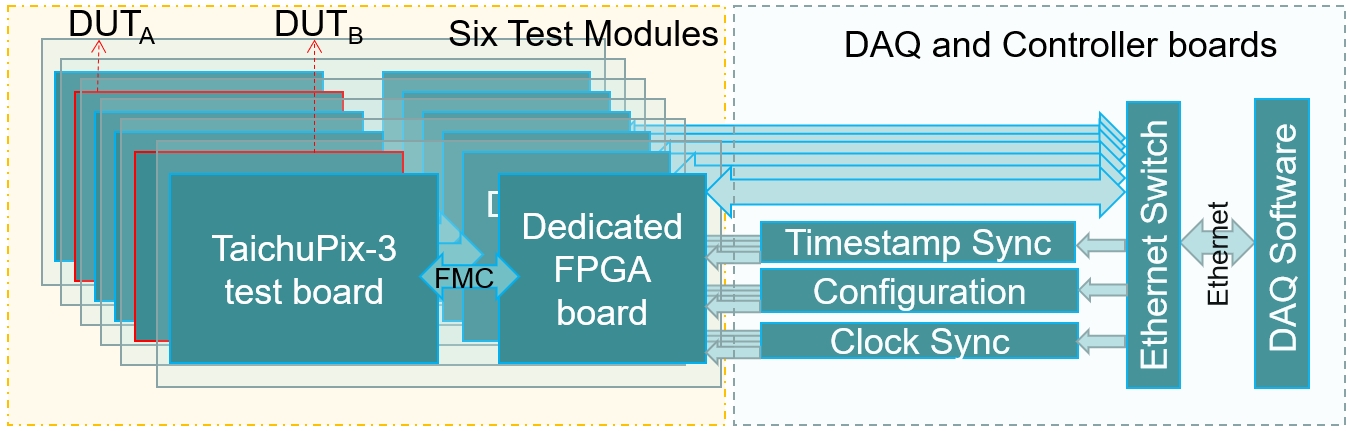}
    \caption{Overall diagram of the detector system.}
    \label{fig:scheme}
\end{figure}

The schematic of the detector system is shown in Fig.~\ref{fig:setup6planes}. The system consists of six single-chip test modules, with two modules designated as DUTs. The distance between adjacent planes is \SI{4}{\cm}. Each module includes a test board with a wire-bonded TaichuPix-3 chip, a dedicated FPGA readout board, a clock controller port, a global configuration port, and a timestamp synchronization port. To reserve space for wire bonding, a window measuring \qtyproduct{1.2x0.9}{\cm} has been cut out from the test board, which accounts for approximately $\qty{26}\%$ of the full chip size. The data within the window is used for offline analysis to minimize the effects of multiple scattering. \\
An overall diagram of the system is presented in Fig.~\ref{fig:scheme}. The diagram can be divided into two main parts: the six test modules and the data acquisition (DAQ) software with controller boards. The dedicated FPGA readout board is directly connected to the test board via the FMC interface, and the readout data is transmitted to a PC through an Ethernet port. A specialized DAQ software has been developed to read out the data from all six channels using a multi-threading method that matches the data rate generated by the beam. To synchronize the data from all six channels, a clock fan-out board (labeled as "Clock Sync" in Fig.~\ref{fig:scheme}) is employed to provide a \SI{20}{\MHz} clock to each module. A frame header is encoded in each run of valid data collection. A configuration board (labeled as "Configuration" in Fig.~\ref{fig:scheme}) is designed to initialize every frame of data, while a timestamp synchronization board (labeled as "Timestamp Sync" in Fig.~\ref{fig:scheme}) is used to calibrate the time delay from each channel.
The TaichuPix-3 chip is powered by a \SI{1.8}{\V} DC supply, with the substrate connected to \SI{0}{\V}. 

\subsection{DUT setup in DESY TB21}
\begin{figure}[ht]
    \begin{center}
    \subfigure[]{\includegraphics[width=0.46\textwidth]{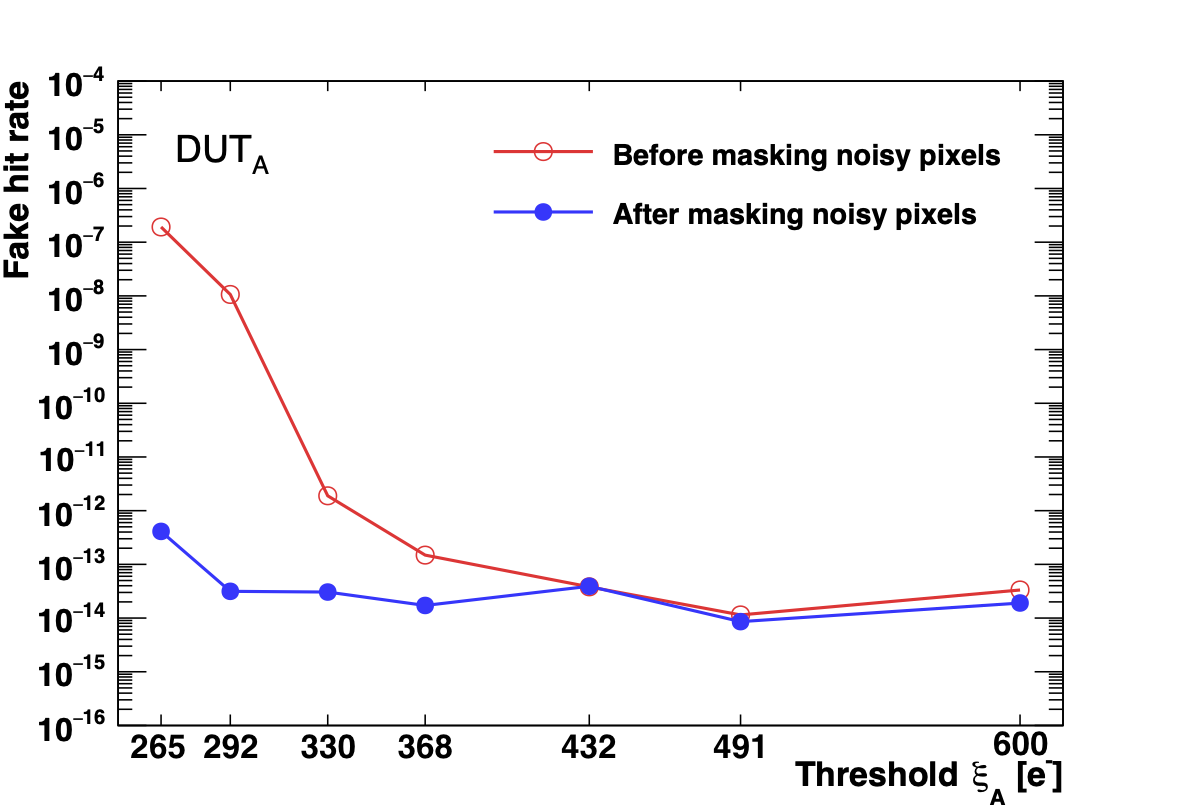}\label{fig:B5_noiserate} }
    \subfigure[]{\includegraphics[width=0.46\textwidth]{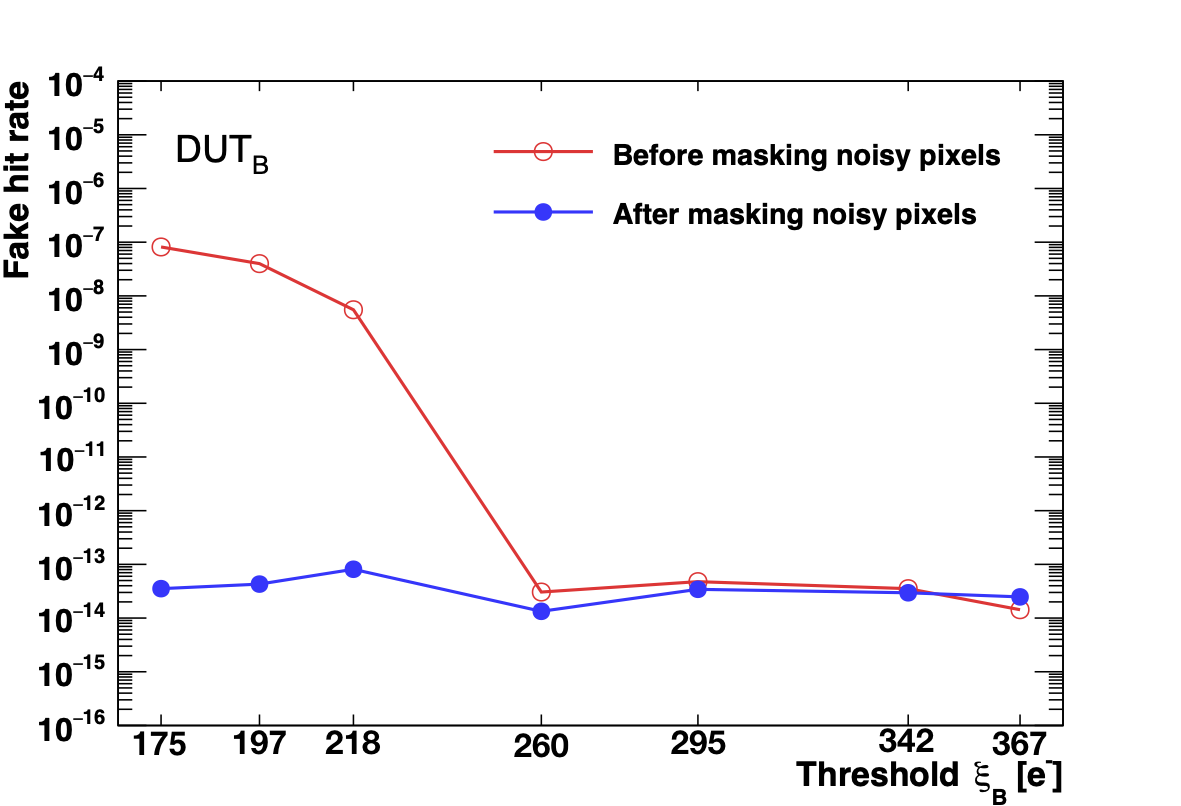}\label{fig:B2_noiserate} }
    \caption{Fake hit rate of  DUT$_{A}$ (a) and DUT$_{B}$(b) as a function of threshold $\xi$. }
    \label{fig:noiserate}
    \end{center}
\end{figure}

\begin{figure}[ht]
    \centering
    \includegraphics[width=1\textwidth]{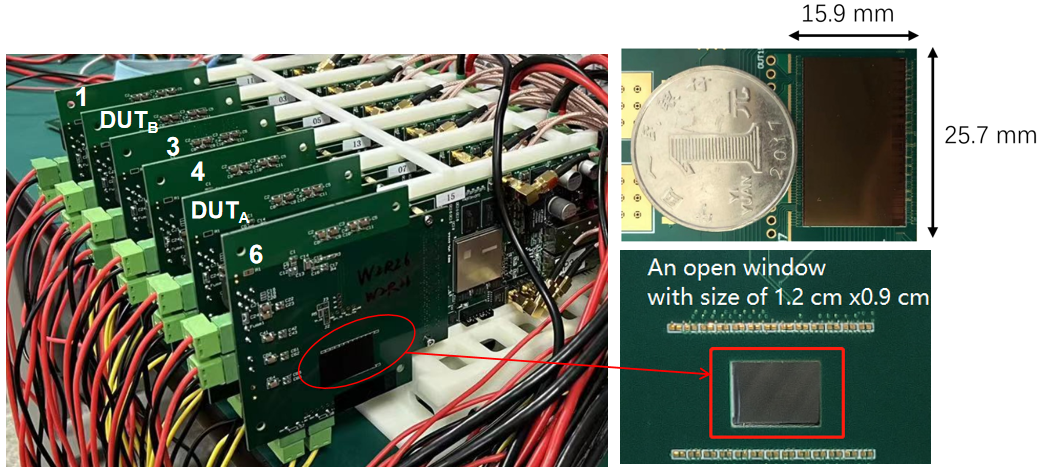}
    \caption{The testing system setup in DESY\uppercase\expandafter{\romannumeral2}.}
    \label{fig:setup6planes}
\end{figure}

The electron-positron synchrotron DESY \uppercase\expandafter{\romannumeral2}~\cite{DIENER:2019265} is capable of providing a particle flux up to $\qty{4}{\kHz\per\square\cm}$ and beam energy is available from 1 to \SI{6}{\GeV}. The experimental setup consists of six planes, each equipped with the same test modules shown in Fig.~\ref{fig:setup6planes}. One test module is configured as a DUT, while the remaining five modules serve as a reference telescope. To compare the performance of two fabrication processes, the fifth plane utilizes TaichuPix-3 with process A (referred to as DUT$_{A}$), while the second plane uses process B (referred to as DUT$_{B}$). The primary beam energy used is \SI{4}{\GeV}, operating at a data rate of $67.4$ Kb$\cdot$\SI{}{\s}$^{-1}$ per plane.

\begin{figure}[ht]
    \centering
    \includegraphics[width=1\textwidth]{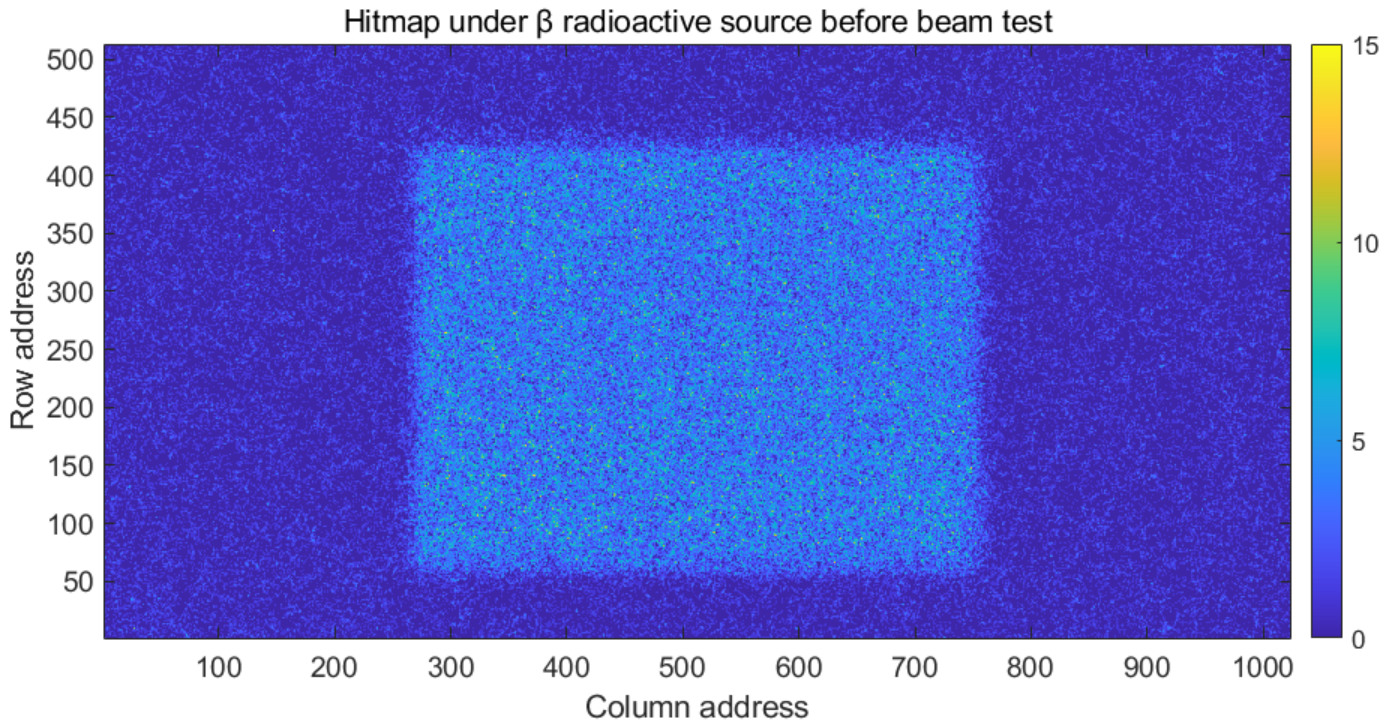}
    \caption{Hitmap under \isotope[90]{Sr} from backside before beam test.}
    \label{fig:HitmapSr}
\end{figure}

Prior to the beam test, each TaichuPix-3 chip undergoes wafer-level selection and is subjected to wire bonding and functional verification through charge injection capacitance within each pixel. A threshold scan is performed to measure the fake hit rate of each chip. As shown in Fig.~\ref{fig:noiserate}, before masking noisy pixels, the maximum fake hit rate is approximately $10^{-7}$ /event/pixel. Since a single on-chip DAC is used to adjust the threshold for all pixels, it is impossible to fine-tune individual pixels. After masking noisy pixels, the fake hit rates are controlled to be no more than {10}$^{-12}$/event/pixel for all measured threshold cases. Additional characterization tests using laser and laboratory radioactive sources have been conducted to evaluate each chip. For example, Fig.~\ref{fig:HitmapSr} illustrates a hitmap obtained under the radioactive source \isotope[90]{Sr} test. The beta source is positioned at the backside of the chip wire-bonded board, and the opened window depicted in Fig.~\ref{fig:setup6planes} appears clear, indicating a higher number of hits compared to the surrounding pixels.

\begin{figure}[ht]
    \centering
    \includegraphics[width=1\textwidth]{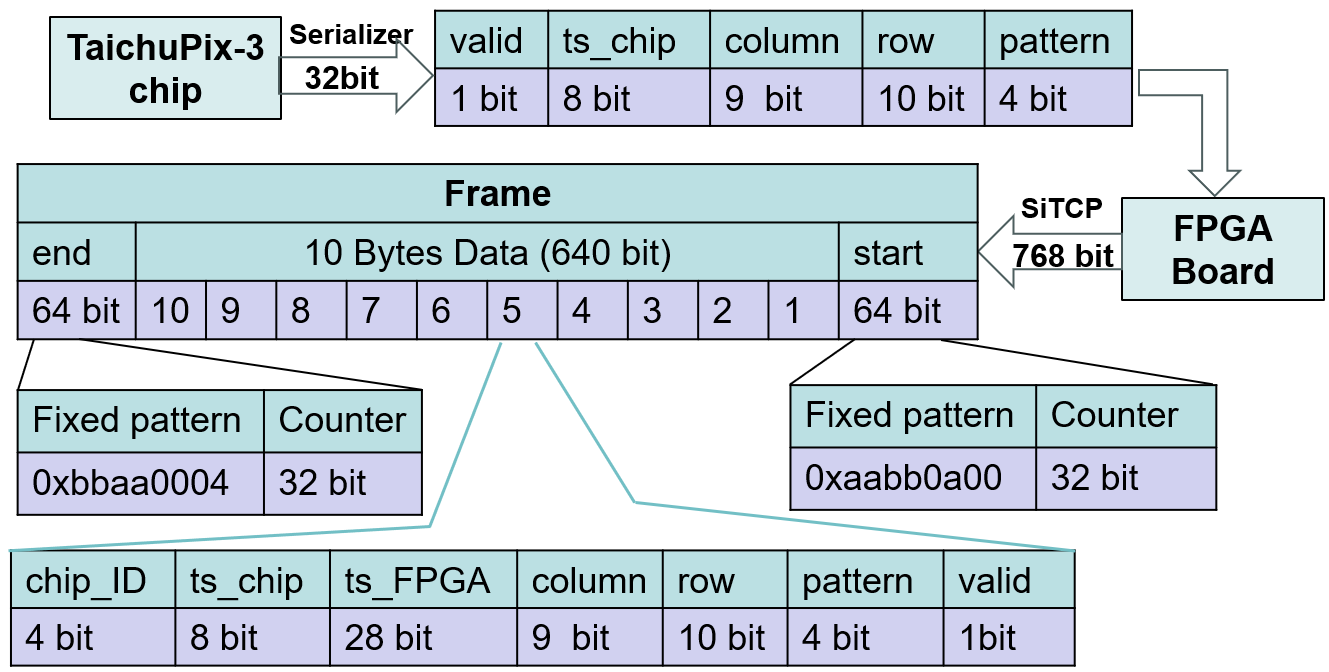}
    \caption{Encoding process for each data acquisition.}
    \label{fig:Encoder}
\end{figure}

Fig.~\ref{fig:Encoder} shows the data readout process from the TaichuPix-3 prototype. The on-chip serializer encodes the data in a 32-bit format, which includes hit information. To align with the DAQ format, ten 32-bit data have been wrapped into a frame of 768 bits by the FPGA board. Each hit data packet consists of a 1-bit data valid flag, a 9-bit column address, a 10-bit row address, an 8-bit on-chip timestamp (ts\_chip), a 28-bit external timestamp (ts\_FPGA), a 4-bit pattern and a 4-bit chip\_ID (used to distinguish the data originating from each chip). A fixed pattern is added at the beginning and the end of each frame, and the DAQ software must identify this fixed pattern before decoding each data segment.

The telescope is operated in triggerless mode and data is taken in the same mode used for fake-hit scans.  For each plane, the capacitance of the charge injection system extracted from the layout is \SI{172}{} aF, as discussed in Ref.~\cite{Zhang:2022rlo}. The threshold voltage is measured using the S-curve method, and the extracted value of the injection capacitance is then used to convert the voltage into an equivalent threshold charge. The measured threshold dispersion encompasses both the variations in the injected signal and the fluctuations in the threshold itself. The threshold-to-ENC (Equivalent Noise Charge) ratio is used to present the extent of the impact of performance noise under the current threshold. In the case of DUT$_A$, the ratio ranged from 4.75 to 8.06, indicating an average threshold range of 265 $e^{-}$ to 600 $e^{-}$. For DUT$_B$, the ratio ranged from 4.18 to 6.38, corresponding to an average threshold range of 175 $e^{-}$ to 367 $e^{-}$. This suggests that a lower threshold will be more affected by ENC.

\FloatBarrier 

\section{Offline analysis and results}\label{sec:Offline analysis and results}
A dedicated offline analysis software has been developed for the analysis of the test beam data. The software primarily includes decoding raw data, clustering, alignment, and track fitting. The alignment of the telescope reference planes and DUT sensors is achieved by minimizing the track residuals using the Millepede algorithm~\cite{BLOBEL20065}. The track fitting is performed using the General Broken Line (GBL) method~\cite{KLEINWORT2012107}, which takes into account the effects of multiple scattering.
\subsection{Cluster size}
\begin{figure}[ht]
    \begin{center}
\subfigure[]{\includegraphics[width=0.46\textwidth]{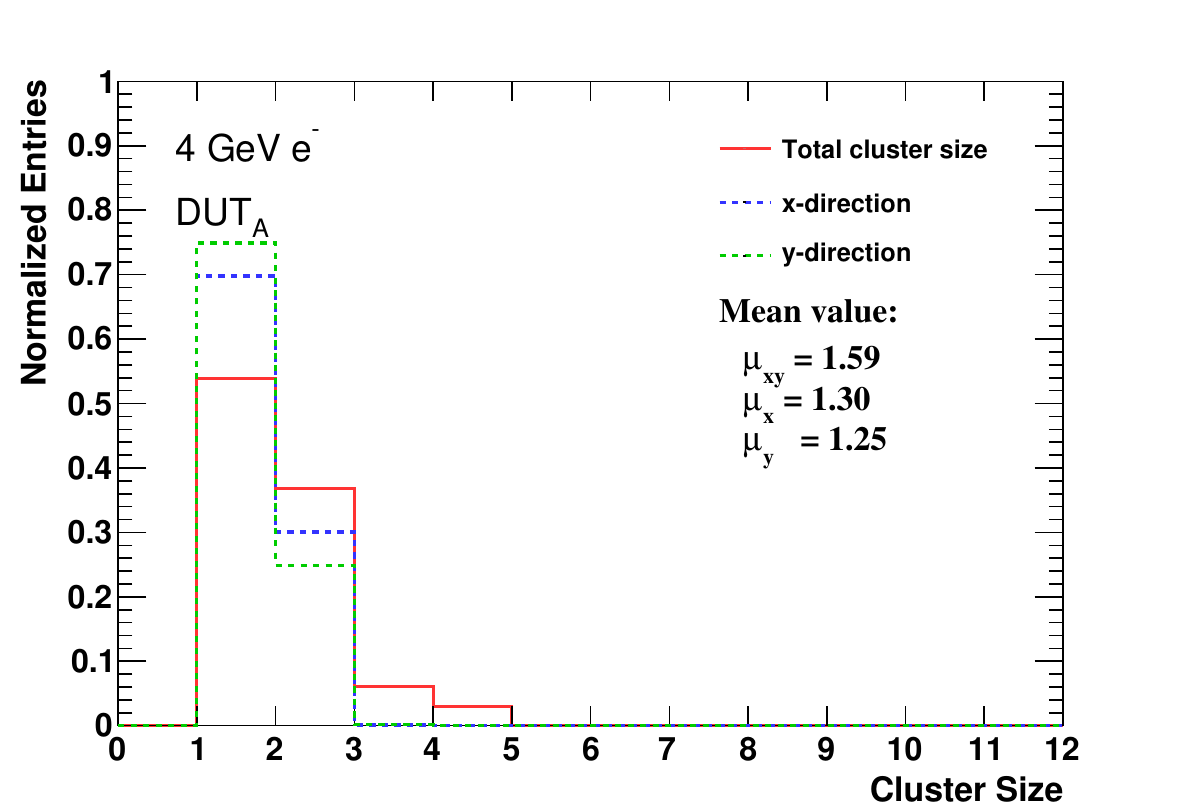}\label{fig:standard_dis} }
    \subfigure[]{\includegraphics[width=0.46\textwidth]{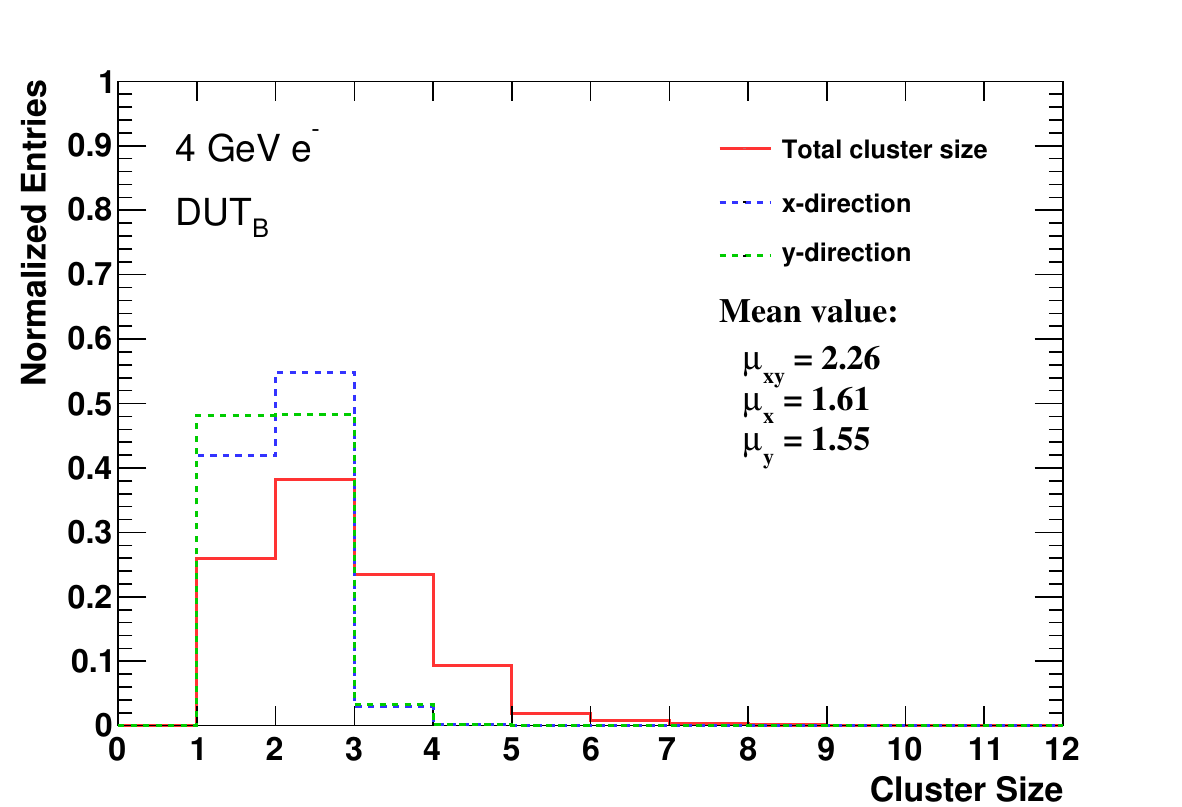}\label{fig:modify_dis} }
    \caption{The cluster size distribution for DUT$_{A}$ with $\xi_{A} = 265$ $e^{-}$ (a) and DUT$_{B}$ with $\xi_{B} = 175$ $e^{-}$ (b), shown in the total cluster size and the cluster size projected onto the $x$-direction (row direction of the sensor) and $y$-direction (column direction of the sensor).}
    \label{fig:ClusterSize_distribution}
    \end{center}
\end{figure}

When beam particles traverse the detector, electron-hole pairs are generated along their trajectory in the epitaxial layer of the sensor. Due to the diffusion of the charge carrier, the charge is shared among neighboring pixels. The clustering is performed by grouping together neighboring fired pixels with signals above a certain threshold $\xi$. The position of the cluster is determined as the geometric center of these grouped pixels. Fig.~\ref{fig:ClusterSize_distribution} illustrates the distribution of cluster size for DUT$_{A}$ and DUT$_{B}$, with a minimum threshold at $\xi_{A} = 265$ $e^{-}$ and $\xi_{B} = 175$ $e^{-}$. The peak cluster size for DUT$_{A}$ is one pixel, while for DUT$_{B}$ it is approximately two pixels.

\begin{figure}[ht]
    \begin{center}
    \subfigure[]{\includegraphics[width=0.46\textwidth]{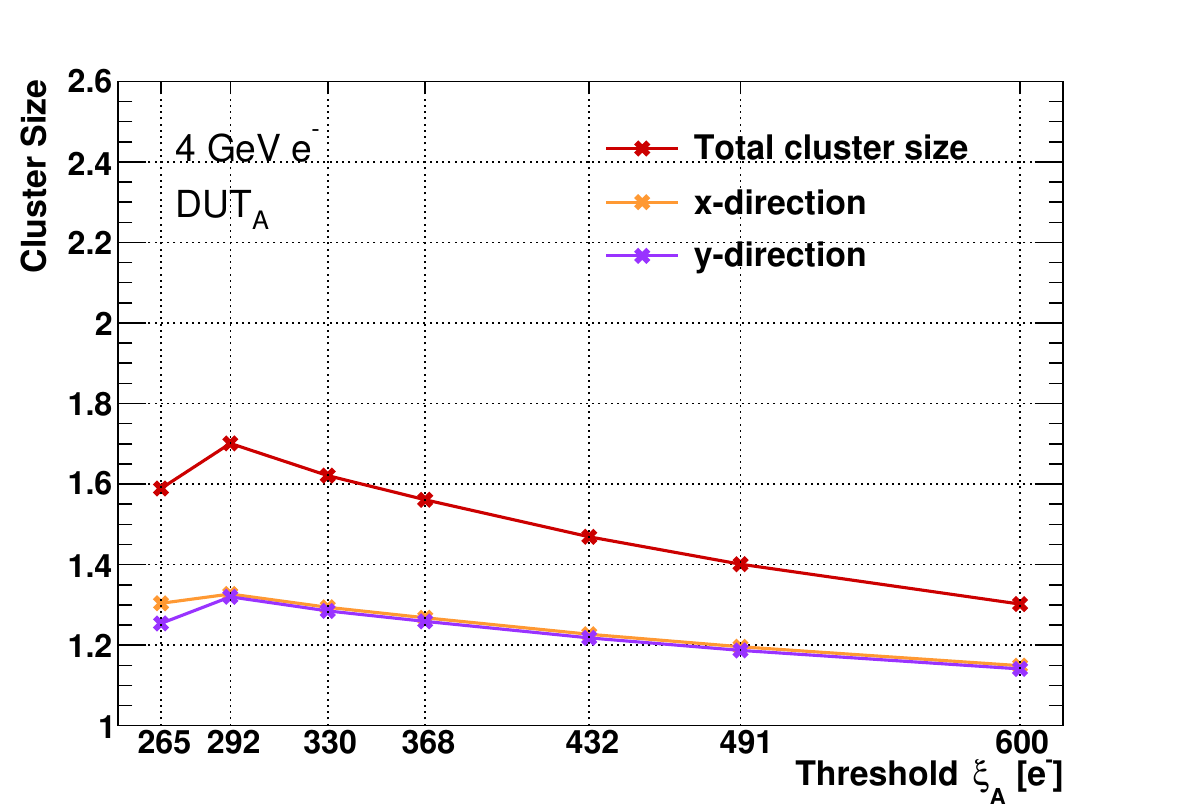}\label{fig:standard_thr} }
    \subfigure[]{\includegraphics[width=0.46\textwidth]{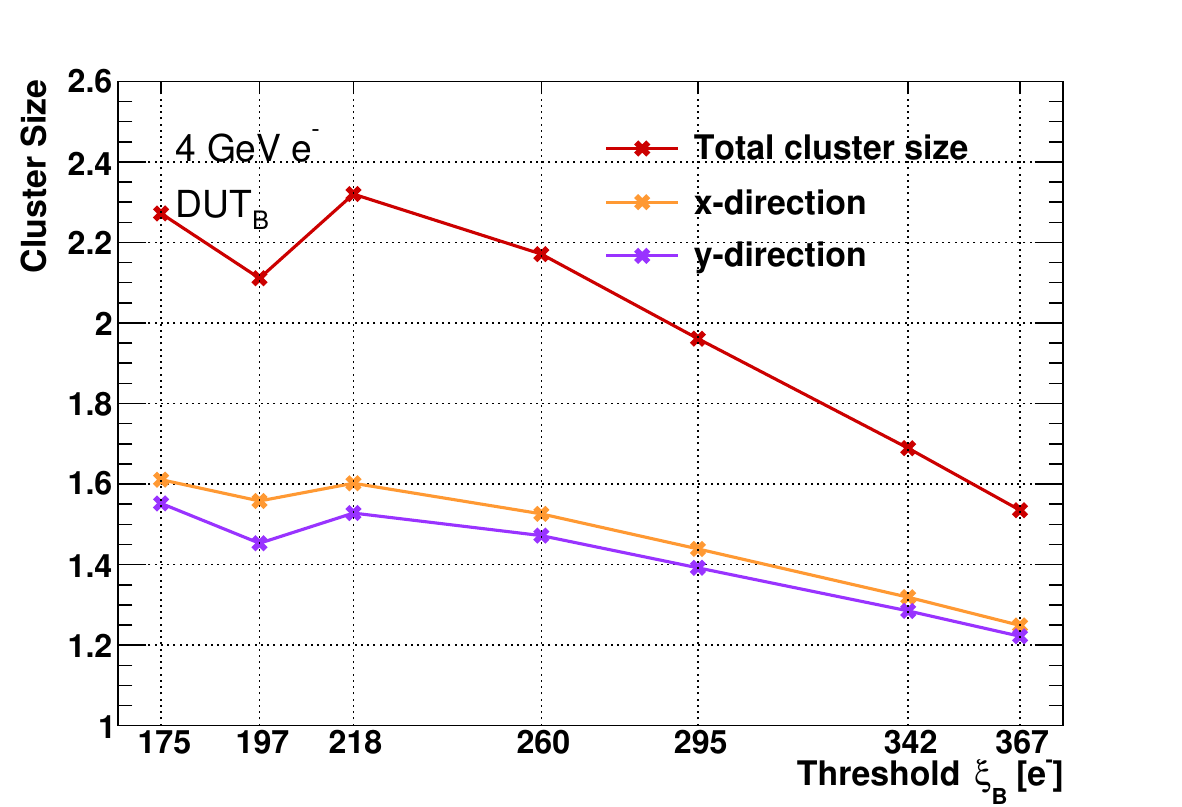}\label{fig:modify_thr} }
    \caption{Average cluster size of DUT$_{A}$ (a) and DUT$_{B}$ (b) as a function of threshold $\xi$, shown in the total cluster size and the cluster size projected onto the $x$-direction and $y$-direction.}
    \label{fig:ClusterSize_threshold}
    \end{center}
\end{figure}

The threshold is a crucial parameter for studying detector performance. A higher threshold reduces the number of fired pixels, resulting in a smaller cluster size. Fig.~\ref{fig:ClusterSize_threshold} displays the average cluster size for DUT$_{A}$ and DUT$_{B}$. As expected, when $\xi_{A} >= 292$ $e^{-}$ and $\xi_{B} >= 218$ $e^{-}$, the cluster size decreases as the threshold increases. However, an abnormal trend is observed when $\xi_{A} < 292$ $e^{-}$ and $\xi_{B} < 218$ $e^{-}$. This is because, at low thresholds, a significant amount of noise leads to deviation in the distribution of cluster size, especially in $y$-direction due to column-based readout. Overall, DUT$_{B}$ shows a larger average cluster size than DUT$_{A}$, indicating less charge-sharing effects on DUT$_{A}$ due to an additional deep N-layer mask, as reported in Ref.~\cite{W:2017technogy}.

\subsection{Alignment}
\begin{figure}[ht]
    \begin{center}
    \subfigure[]{\includegraphics[width=0.46\textwidth]{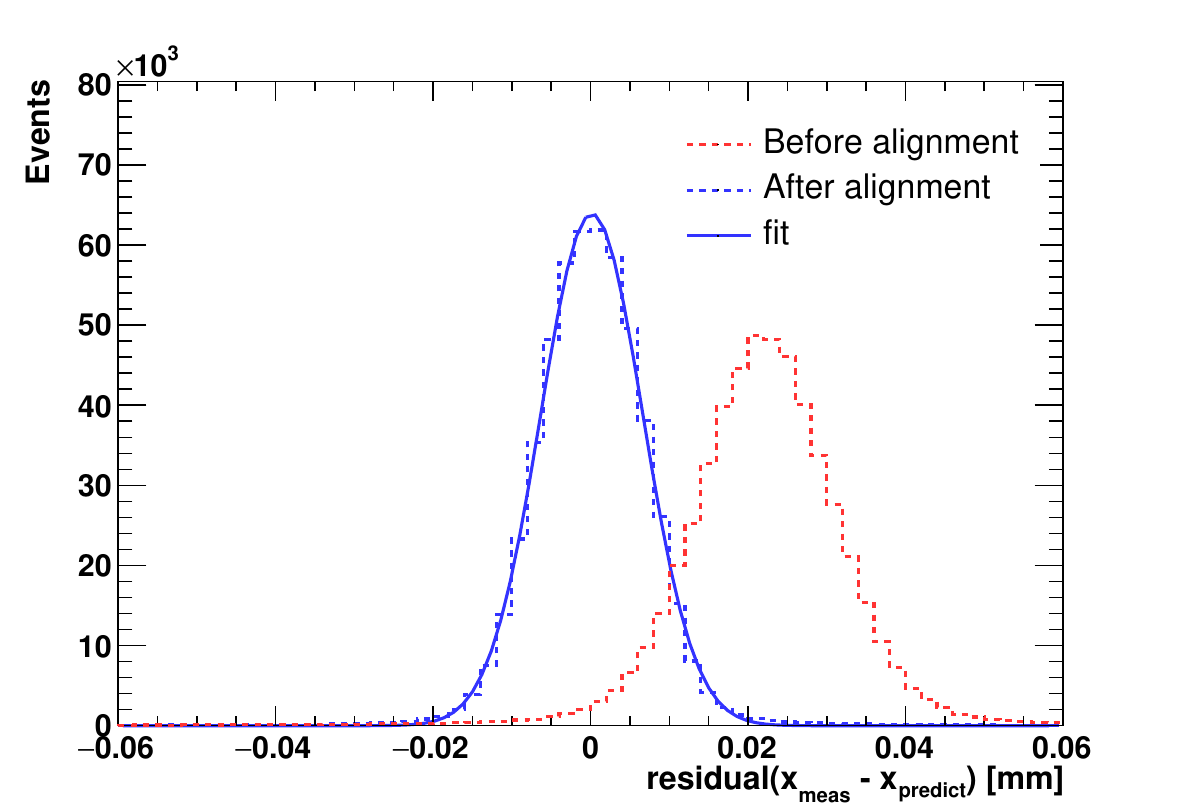}\label{fig:befalignx} }
    \subfigure[]{\includegraphics[width=0.46\textwidth]{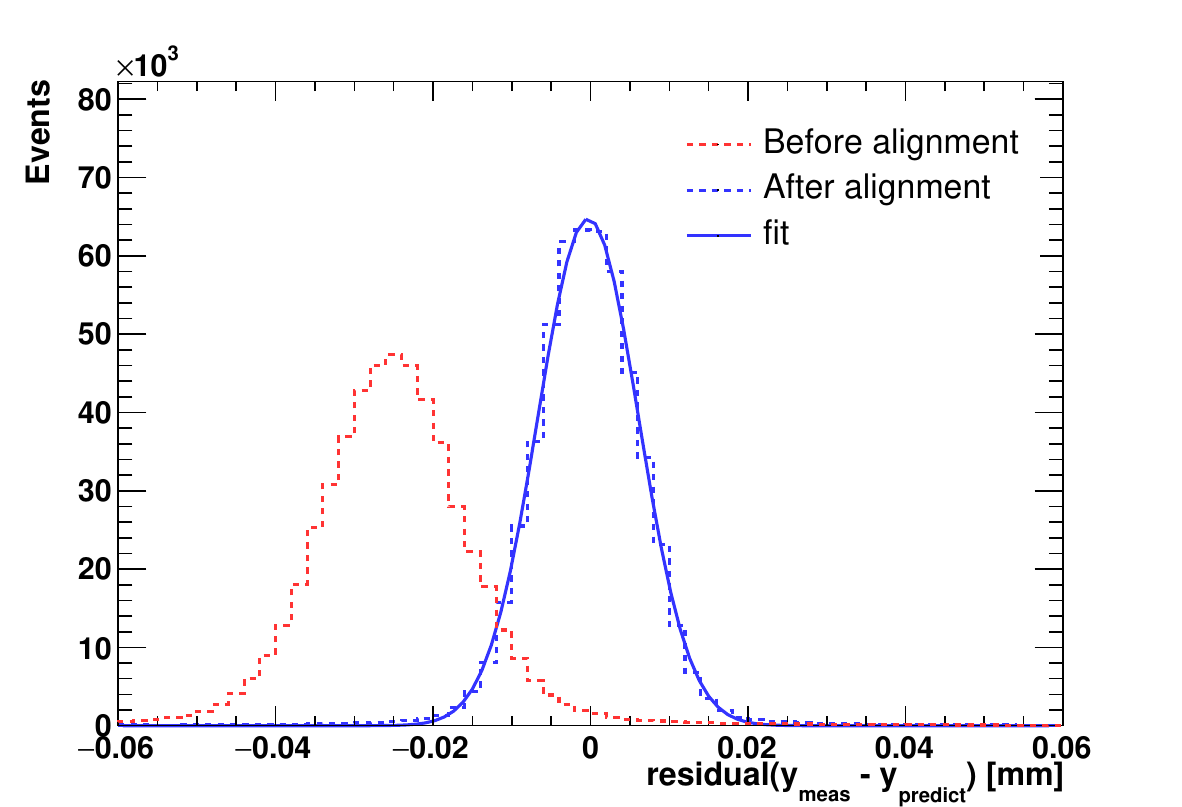}\label{fig:befaligny} }
    \caption{The residual distribution of an example DUT in $x$-direction (a) and $y$-direction (b). The residual distribution after alignment is fitted with a Gaussian function without applying a track quality $\chi ^2$ cut.}
    \label{fig:befalign}
    \end{center}
\end{figure}

The differences between the assumed detector geometry and the actually installed detector geometry can introduce bias in the measured hit positions, leading to potential degradation of the spatial resolution. Hence, alignment plays a vital role in the analysis of beam test data. The alignment parameters, including three offsets and three rotations for each plane with respect to a global common reference system, are determined iteratively using a general $\chi ^2$ minimization procedure. The $\chi^2$ function is given by Eq.~\ref{eq1}.
\begin{equation}
\label{eq1}
\chi^2 = \sum_{j\in tracks}\sum_{i\in hits}\vec{r}^{T}_{ij}(g, l_j)V_{ij}^{-1}\vec{r}_{ij}(g, l_{j})
\end{equation}
where $V_{ij}$ represents the covariance matrix of the measurements, and $\vec{r}_{ij}$ represents the residual between the measured hit position on the DUT and the hit position predicted by the track model, which is related to the local parameters ($l$) of each track $j$ and the global parameters ($g$) i.e., the alignment parameters. The solution of minimizing Eq.~\ref{eq1} is calculated by the Millepede~\cite{BLOBEL20065} program which can simplify this equation by matrix reduction, giving the solution of the six alignment parameters. Then, the alignment parameters are applied to the geometry of the detector. Fig.~\ref{fig:befalign} illustrates the residual distributions of an example DUT in the $x$- direction and $y$-direction before and after alignment.
\subsection{Spatial resolution studies}
The estimation of the intrinsic resolution of DUT and the track resolution is discussed in this section. The intrinsic resolution is estimated by the iterative pull method described in Ref.~\cite{article}. The formalism of GBL is used for track fitting, which takes into account the effect of multiple scattering from \SI{150}{\um} of silicon per sensor and air between adjacent planes. The standard deviation $\theta_{0}$ of scattering angular distribution is calculated using the Highland model~\cite{HIGHLAND1975497} written as:

\begin{equation}
\label{eq2}
\theta_{0} = \frac{13.6 MeV}{\beta c p}Z \sqrt{\varepsilon}(1+0.038 ln(\varepsilon))
\end{equation}

where $p$ and Z are the momentum and the charge number of the beam particle, and $\varepsilon$ represents the material budget that the particle traverses. After determining the intrinsic resolution of each plane, the track resolution at the DUT position $z$ is derived from the covariance matrix using the formalism of GBL.

\subsubsection{Intrinsic resolution estimation}

The intrinsic resolution $\sigma_{int}(z)$ of the DUT plane at position $z$ can be determined by evaluating the width of unbiased $r_{unb}(z)$ residual distribution, as demonstrated in Eq.~\ref{eq3}.
\begin{equation}
\label{eq3}
r_{unb}^2(z) = \sigma_{int}^2(z)+\sigma_{trc, unb}^2(z)
\end{equation}

$\sigma_{trc, unb}$ represents the unbiased and biased track resolution. Unbiased tracks exclude the measured hits of the DUT, in contrast to biased tracks which include the hits of the DUT in the track fits.

According to Eq.~\ref{eq3}, the unbiased pull of a track can be defined as follows:
\begin{equation}
\label{eq5}
p_{unb} = \frac{r_{unb}}{\sqrt{\sigma_{int}^2+\sigma_{trc, unb}^2}}
\end{equation}

Following the methodology outlined in Ref.~\cite{article}, if the $\theta_0$ of the scattering angular distribution and the intrinsic resolution $\sigma_{int}(z)$ of each sensor are accurately described, the pull distribution $p_{unb}$ of each plane will satisfy $p_{unb} \sim N(0, 1)$. Each plane is known to be a \SI{150}{\um}  of silicon. Initially, an assumed intrinsic resolution of $\hat{\sigma}_{int}(z)$ is input into GBL for every plane. If the standard deviation of $p_{unb}$ deviates from one, the input $\hat{\sigma}_{int}(z)$ is iterated updated until $p_{unb} \sim N(0, 1)$ for each plane. The intrinsic resolution of each plane is determined simultaneously in the iterative process.

The estimation of $\sigma_{int}(z)$ is affected by systematic uncertainties stemming from various sources, including the spread of beam energy (\SI{5}{\%}), the prediction accuracy of the $\theta_0$ (\SI{11}{\%}), and the chosen fit range on the pull distribution. The nominal fit range on the pull distribution is $ \pm 2$. Each individual source of uncertainty is varied by $\pm 1$ $\sigma$ in the track fit and independently propagated to $\sigma_{int}(z)$ using the same iterative method with the nominal value. The final systematic uncertainty is obtained by summing the individual uncertainties in quadrature. The statistical uncertainty is negligible. The estimated intrinsic resolution for six planes, based on the telescope threshold configuration, is shown in Table.~\ref{tab1}. \par

\begin{table}[t]
\caption{Estimated intrinsic resolution for six planes with the listed telescope configuration. The uncertainty is the total systematic uncertainty.}
\centering
\label{tab1}
\begin{tabular}{ccccc}
\toprule
\multirow{2}*{Plane number} & \multirow{2}*{Process} &\multirow{2}*{Threshold [$e^{-}$]}& \multicolumn{2}{c}{intrinsic resolution} \\
\cline{4-5}
& & & $x$-direction & $y$-direction\\
\midrule
Plane 0&A&287& 4.82 $\pm$ 0.14 & 4.88 $\pm$ 0.15\\

Plane 1&B&260& 4.65 $\pm$ 0.12 & 4.79 $\pm$ 0.11\\
Plane 2&A&298& 4.90 $\pm$ 0.15 & 4.95 $\pm$ 0.15\\

Plane 3&B&281& 4.86 $\pm$ 0.17 & 4.97 $\pm$ 0.17\\
Plane 4&A&292& 4.84 $\pm$ 0.10 & 4.90 $\pm$ 0.10\\

Plane 5&A&294& 4.88 $\pm$ 0.14 & 4.93 $\pm$ 0.15\\
\bottomrule
\end{tabular}
\end{table}

\begin{figure}[ht]
    \begin{center}
    \subfigure[]{\includegraphics[width=0.46\textwidth]{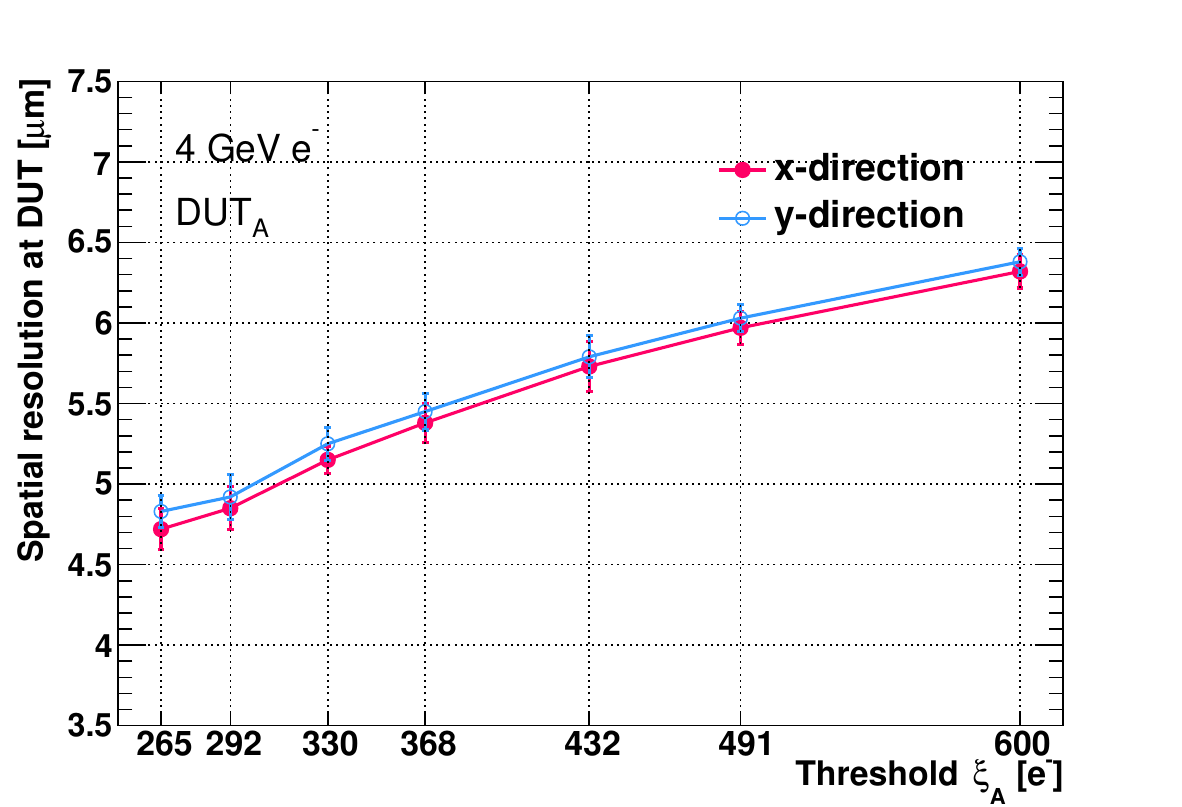}\label{fig:res_thr_standard} }
    \subfigure[]{\includegraphics[width=0.46\textwidth]{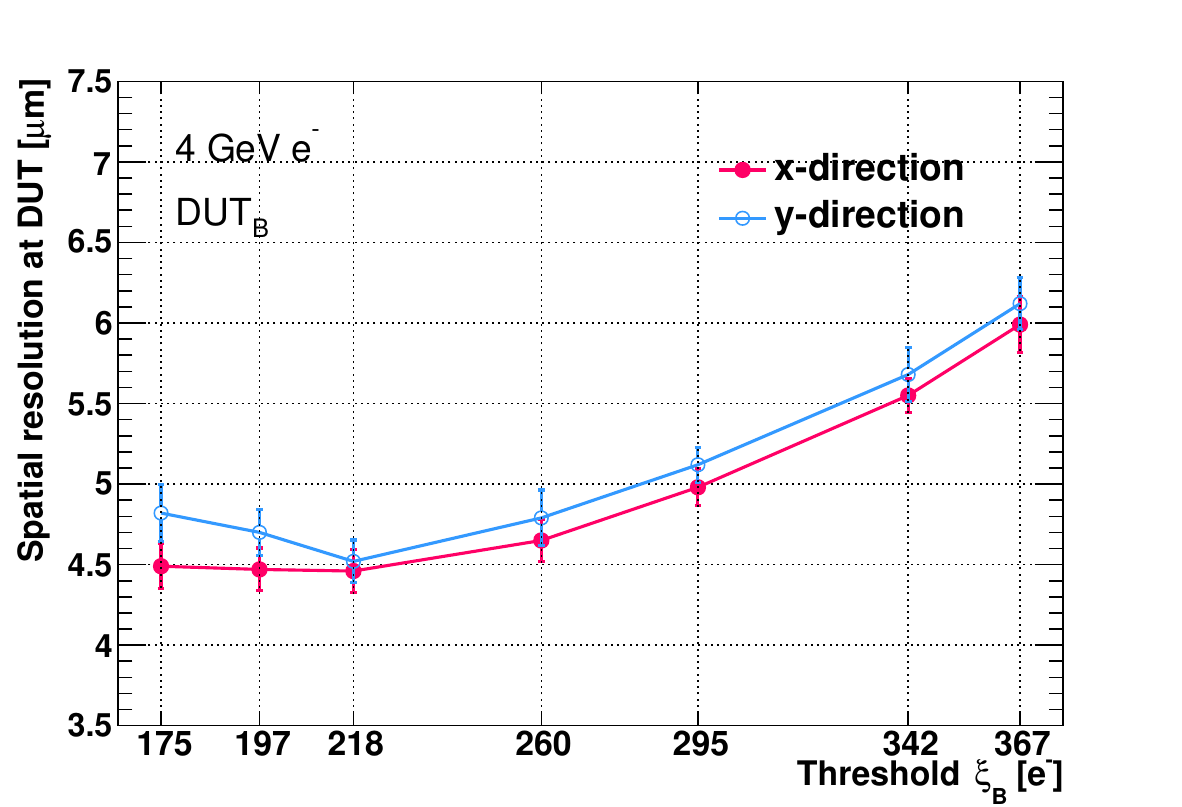}\label{fig:res_thr_modify}}
    \caption{Spatial resolution as a function of threshold for DUT$_{A}$(a) and DUT$_{B}$(b) in the $x$-direction and $y$-direction. The error bars represent the total systematic uncertainty.}
    \label{fig:Resolution_Threshold}
    \end{center}
\end{figure}

The threshold not only affects the cluster size but also impacts the intrinsic resolution. In general, a higher threshold leads to a smaller cluster size, which introduces a bias in estimating the actual hitting position and ultimately worsens the intrinsic resolution. As depicted in Fig.~\ref{fig:Resolution_Threshold}, for the two DUTs, increasing the threshold results in a deterioration of the intrinsic resolution. However, for DUT$_{B}$, a worse resolution is also observed when the threshold is lower than $\xi_B = 218$ $e^{-}$, which can be attributed to the increased noise at the lower thresholds. The best resolution for DUT$_{A}$ is 4.72 $\pm$ 0.13 (syst.)~\SI{}{\um} in the $x$-direciton, 4.83 $\pm$ 0.10 (syst.)~\SI{}{\um} in the $y$-direciton when $\xi_B = 265$ $e^{-}$. For DUT$_{B}$, the best resolution is 4.46 $\pm$ 0.13 (syst.)~\SI{}{\um} int the $x$-direciton, 4.52 $\pm$ 0.13 (syst.)~\SI{}{\um} int the $y$-direciton when $\xi_B = 218$ $e^{-}$ .

\subsubsection{Track resolution on DUT }

\begin{figure}[ht]
    \begin{center}
    \subfigure[]{\includegraphics[width=0.46\textwidth]{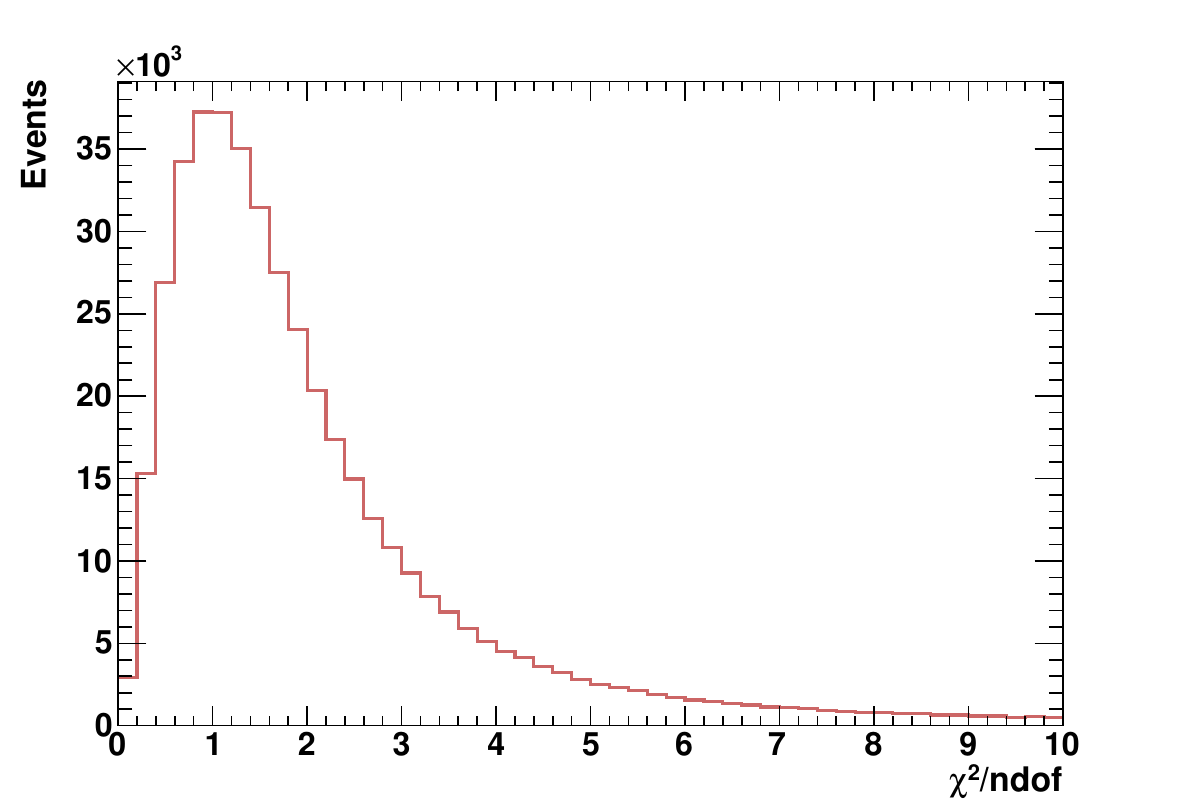}\label{fig:chi2}}
    \subfigure[]{\includegraphics[width=0.46\textwidth]{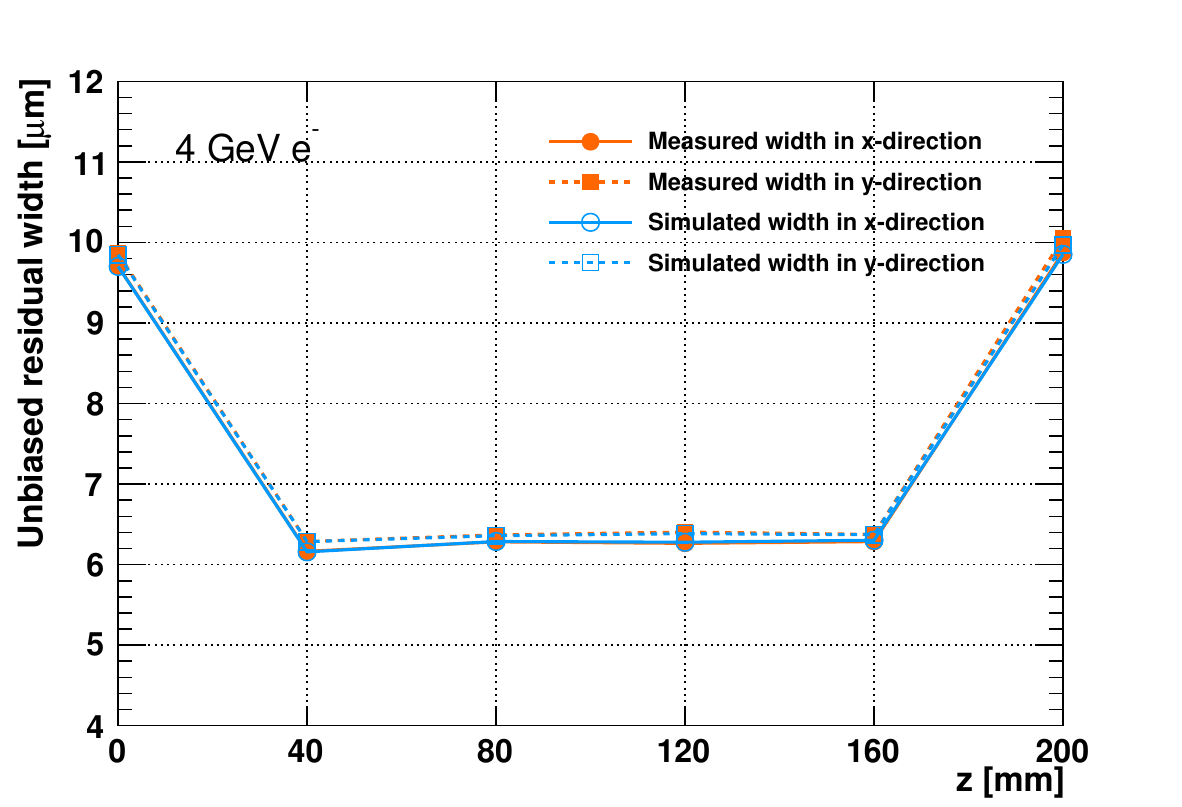}\label{fig:position}}
    \caption{(a) Distribution of the $\chi^2$ per degree of freedom. (b) Measured and simulated unbiased residual width in the $x$-direciton and $y$-direciton for six planes.}
    \label{fig:position_resolution}
    \end{center}
\end{figure}

After determining the intrinsic resolutions of the telescope planes, the track resolution at the $z_{DUT_{A}}$ and $z_{DUT_{B}}$ can be extracted from the covariance matrix of the track using the GBL formalism. Fig.~\ref{fig:position_resolution}\subref{fig:chi2} illustrates the total $\chi^2$ per degree of freedom of the track. A track quality cut of $\chi^{2}/ndof < 100$ is applied to the track fitting. The measured unbiased residual widths of each plane in the $x$-direciton and $y$-direciton are shown in Fig.~\ref{fig:position_resolution}\subref{fig:position}. It can be observed that the inner planes exhibit the minimum width, while the unbiased residual width increases towards the outer planes. Furthermore, it can be inferred from Eq.~\ref{eq3} that the track resolution at the outer planes is worse than that at the inner planes.

The track resolution is extracted from the covariance matrix of the track parameters using the GBL track fitting. The intrinsic resolution obtained through the iterative pull method is used as the input for the measurement uncertainty of the GBL fitting, taking into account the material budget of all planes and the multiple scattering from the air between adjacent planes. The systematic uncertainties of the track resolution come from the \SI{5}{\%} uncertainty in the beam energy and the \SI{11}{\%} uncertainty in the calculation of the standard deviation, $\theta_0$, of the scattering angle. Therefore, the track resolutions at $z_{DUT_{A}}$ and $z_{DUT_{B}}$ are as follows:

\begin{equation}
\label{eq6}
\begin{split}
\sigma_{trc, x}(z_{DUT_{A}}) = 4.010 \pm 0.086~ (syst.)~ \SI{}{\um} \\
\sigma_{trc, y}(z_{DUT_{A}}) = 4.049 \pm 0.083~ (syst.)~ \SI{}{\um} \\
\sigma_{trc, x}(z_{DUT_{B}}) = 3.975 \pm 0.088~ (syst.)~ \SI{}{\um} \\
\sigma_{trc, y}(z_{DUT_{B}}) = 4.004 \pm 0.083~ (syst.)~ \SI{}{\um}
\end{split}
\end{equation}

\subsection{Detection efficiency}
The detection efficiency $\epsilon$ is defined as the ratio of tracks that can match the hits on the DUT plane within a distance $d$ around the predicted hits to all tracks. The distance $d$ is set to be $100 $ $\mu m$ to avoid inefficiencies caused by poorly reconstructed tracks. Therefore, the detection efficiency can be calculated using the following formula:
\begin{equation}
\label{eq7}
\epsilon = \frac{N_{|x_{meas}, y_{meas}-x_{pre}, y_{pre}| < d}^{matched\;tracks }}{N_{all}^{tracks}}
\end{equation}

\begin{figure}[ht]
    \begin{center}
    \subfigure[]{\includegraphics[width=0.46\textwidth]{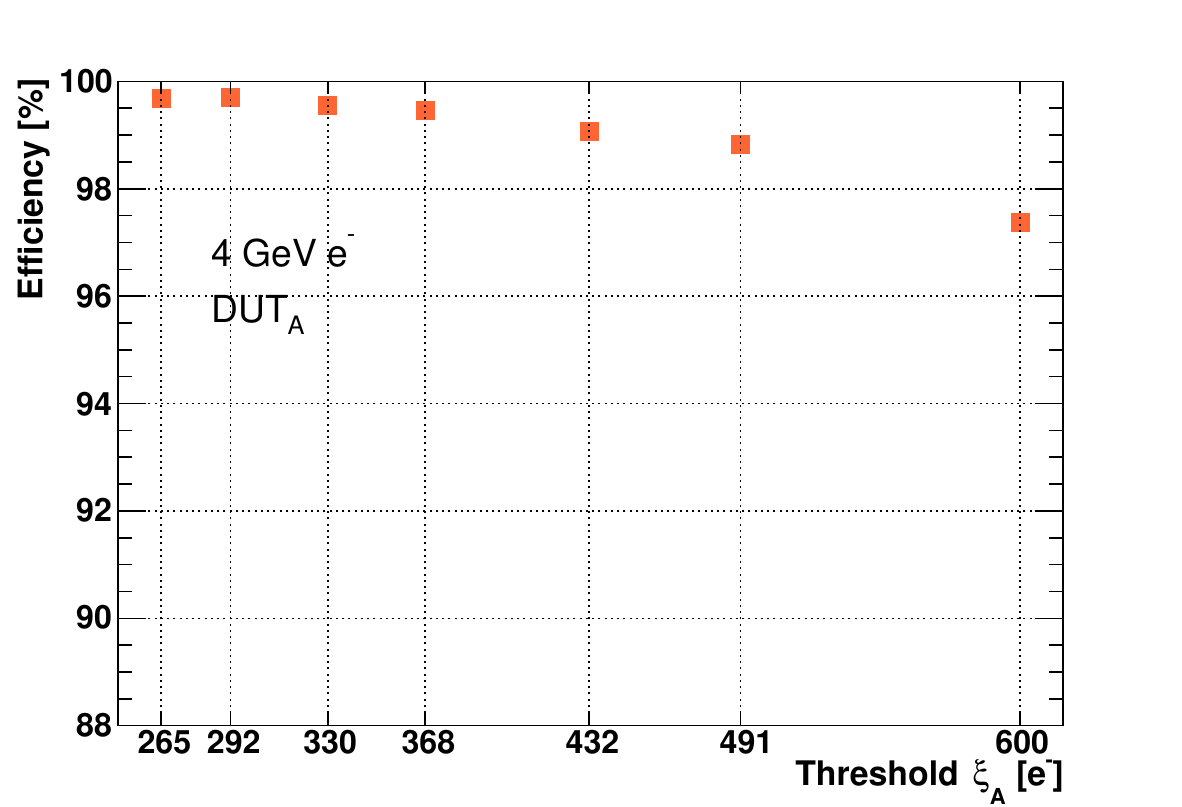}\label{fig:eff_A} }
    \subfigure[]{\includegraphics[width=0.46\textwidth]{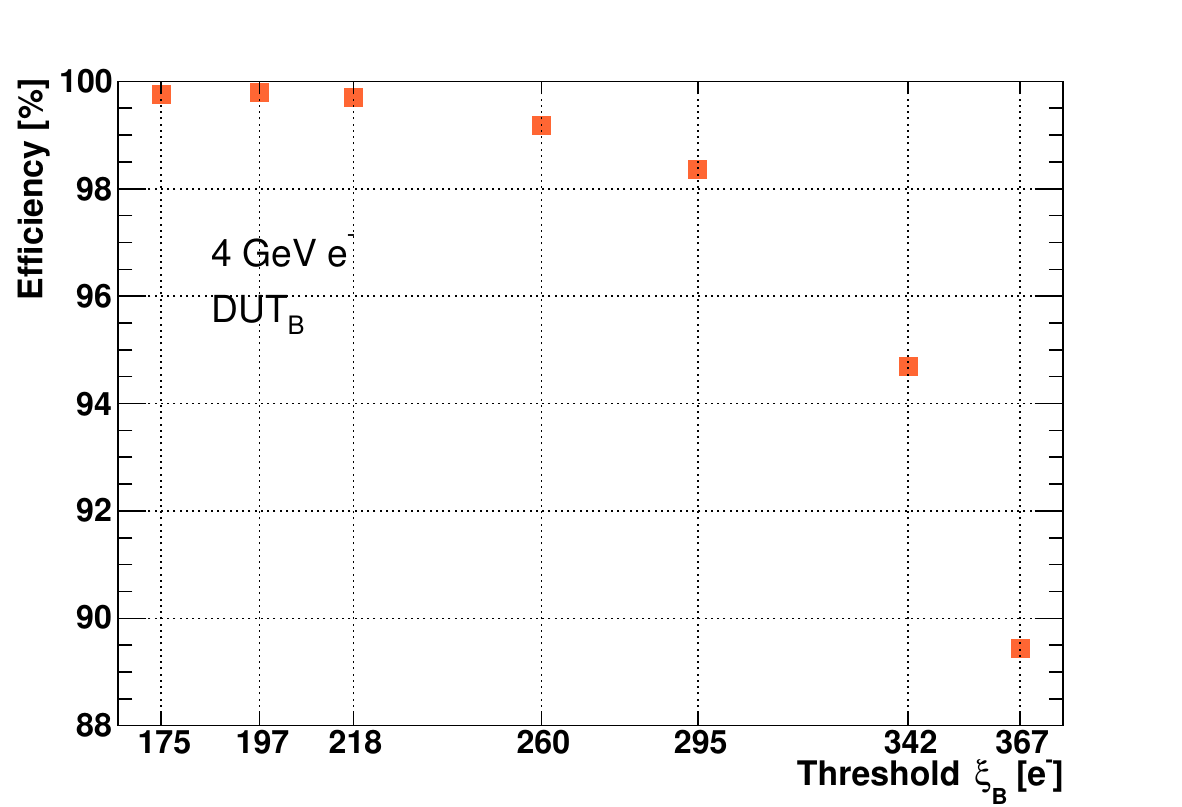}\label{fig:eff_B}}
    \caption{The detection efficiency of DUT$_{A}$ and DUT$_{B}$ as a function of threshold.}
    \label{fig:Eff_Threshold}
    \end{center}
\end{figure}

Fig.~\ref{fig:Eff_Threshold} shows that the efficiencies of DUT$_{A}$ and DUT$_{B}$ decrease as the threshold increases. The maximum detection efficiency is \SI{99.68}{\%} for DUT$_{A}$ and \SI{99.76}{\%} for DUT$_{B}$. However, the efficiency of DUT$_{B}$ drops significantly at high thresholds compared to DUT$_{A}$. This can be explained by the difference in the process of the two DUTs. According to Ref.~\cite{W:2017technogy}, an additional low-dose N-layer is added to DUT$_{A}$ based on DUT$_{B}$. This modification enables a larger depletion of the epitaxial layer and results in a larger charge collection area.

\begin{figure}[ht]
    \begin{center}
    \includegraphics[width=1\textwidth]{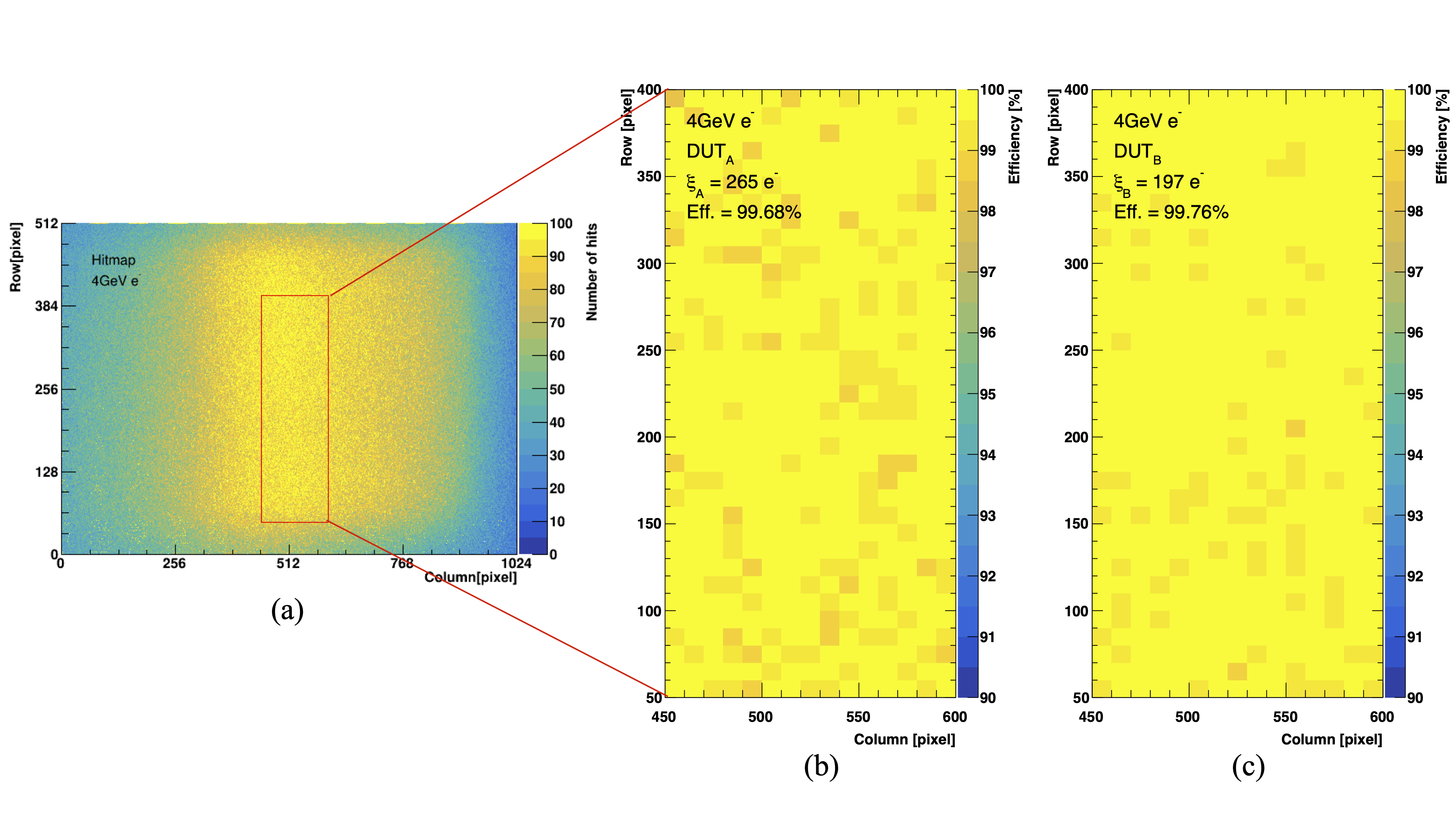}
    \caption{(a) The hitmap of one example DUT under 4 GeV electron beam. The pixels inside the red box are used to calculate the average efficiency of every 10 × 10 pixels. (b) (c) The efficiency map of DUT$_{A}$ and DUT$_{B}$ at the optimal threshold.}
    \label{fig:effmap}
    \end{center}
\end{figure}

The hitmap of one example DUT covering the full pixel area is shown in Fig.~\ref{fig:effmap}(a). The efficiency map of two DUTs is displayed in Fig.~\ref{fig:effmap}(b) and (c), respectively, indicating good uniformity.

\FloatBarrier

\section{Conclusion}\label{sec:conclusion}
A MAPS prototype, TaichuPix-3, was developed for the CEPC vertex detector. A testing system including six detector modules with TaichuPix-3 chips, readout electronics and the DAQ system was set up and tested with a 4 GeV electron beam at DESY \uppercase\expandafter{\romannumeral2} TB21. This work discusses the characterizations of TaichuPix-3 including cluster size, spatial resolution, detection efficiency and the differences in characteristics of the two processes. Additionally, the telescope resolution composed of TaichuPix-3 chips is provided.
The analysis results show that the best spatial resolution is about \SI{4.8}{\um} for DUT$_{A}$ and \SI{4.5}{\um} for DUT$_{B}$ and the detection efficiency can reach \SI{99.5}{\%} for both processes, which achieve the current design goal. The tracking resolution of the telescope composed of TaichuPix-3 chips is about \SI{4.0}{\um} at the inner planes. The baseline vertex detector consisting of six ladders with double-sides TaichuPix-3 chips will be tested at DESY \uppercase\expandafter{\romannumeral2}. The beam test results of TaichuPix-3 are promising for the baseline vertex detector to meet a single point spatial resolution target of better than \SI{5}{\um}.

\section{Acknowledgment}
The research was supported and financed in large part by the National Key
Research and Development Program of China under Grant No. 2018YFA0404302
from the Ministry of Science and Technology. The measurements leading to these results have been performed at the Test Beam Facility at DESY Hamburg (Germany), a member of the Helmholtz Association (HGF). Additional support was provided by the
Youth Scientist Fund from the National Natural Science Foundation of China under
Grant No. 12205313, China Postdoctoral Science Foundation under Grant No.
2021M693182 and State Key Laboratory of Particle Detection and Electronics under Grant No. SKLPDE-ZZ-202315.

\FloatBarrier


\begin{thebibliography}{10}
\expandafter\ifx\csname url\endcsname\relax
  \def\url#1{\texttt{#1}}\fi
\expandafter\ifx\csname urlprefix\endcsname\relax\def\urlprefix{URL }\fi
\expandafter\ifx\csname href\endcsname\relax
  \def\href#1#2{#2} \def\path#1{#1}\fi

\bibitem{CEPC-SPPC-accelerator}
{The CEPC Study Group}, {CEPC Conceptual Design Report, in: Volume I – Accelerator} (9 2018).
\newblock \href {https://doi.org/https://doi.org/10.48550/arXiv.1809.00285} {\path{doi:https://doi.org/10.48550/arXiv.1809.00285}}.

\bibitem{CEPC-SPPC-detector}
{The CEPC Study Group}, {CEPC Conceptual Design Report: Volume 2 - Physics \& Detector} (11 2018).
\newblock \href {https://doi.org/https://doi.org/10.48550/arXiv.1811.10545} {\path{doi:https://doi.org/10.48550/arXiv.1811.10545}}.

\bibitem{Wei:2019wbr}
X.~Wei, et~al., {High data-rate readout logic design of a 512 \texttimes{} 1024 pixel array dedicated for CEPC vertex detector}, JINST 14~(12) (2019) C12012.
\newblock \href {https://doi.org/10.1088/1748-0221/14/12/C12012} {\path{doi:10.1088/1748-0221/14/12/C12012}}.

\bibitem{Wu:2021mju}
T.~Wu, et~al., {The TaichuPix1: a monolithic active pixel sensor with fast in-pixel readout electronics for the CEPC vertex detector}, JINST 16~(09) (2021) P09020.
\newblock \href {https://doi.org/10.1088/1748-0221/16/09/P09020} {\path{doi:10.1088/1748-0221/16/09/P09020}}.

\bibitem{Zhang:2022rlo}
Y.~Zhang, et~al., {Development of a CMOS pixel sensor prototype for the high hit rate CEPC vertex detector}, Nucl. Instrum. Meth. A 1042 (2022) 167442.
\newblock \href {https://doi.org/10.1016/j.nima.2022.167442} {\path{doi:10.1016/j.nima.2022.167442}}.

\bibitem{DIENER:2019265}
R.~Diener, J.~Dreyling-Eschweiler, H.~Ehrlichmann, I.~Gregor, U.~Kötz, U.~Krämer, N.~Meyners, N.~Potylitsina-Kube, A.~Schütz, P.~Schütze, M.~Stanitzki, {The DESY II test beam facility}, Nucl. Instrum. Meth. A 922 (2019) 265--286.
\newblock \href {https://doi.org/https://doi.org/10.1016/j.nima.2018.11.133} {\path{doi:https://doi.org/10.1016/j.nima.2018.11.133}}.

\bibitem{W:2017technogy}
W.~Snoeys, et~al., {A process modification for CMOS monolithic active pixel sensors for enhanced depletion, timing performance and radiation tolerance}, Nucl. Instrum. Meth. A 871 (2017) 90--96.
\newblock \href {https://doi.org/http://dx.doi.org/10.1016/j.nima.2017.07.046} {\path{doi:http://dx.doi.org/10.1016/j.nima.2017.07.046}}.

\bibitem{BLOBEL20065}
V.~Blobel, \href{https://www.sciencedirect.com/science/article/pii/S0168900206007984}{Software alignment for tracking detectors}, Nuclear Instruments and Methods in Physics Research Section A: Accelerators, Spectrometers, Detectors and Associated Equipment 566~(1) (2006) 5--13, tIME 2005.
\newblock \href {https://doi.org/https://doi.org/10.1016/j.nima.2006.05.157} {\path{doi:https://doi.org/10.1016/j.nima.2006.05.157}}.
\newline\urlprefix\url{https://www.sciencedirect.com/science/article/pii/S0168900206007984}

\bibitem{KLEINWORT2012107}
C.~Kleinwort, \href{https://www.sciencedirect.com/science/article/pii/S0168900212000642}{General broken lines as advanced track fitting method}, Nuclear Instruments and Methods in Physics Research Section A: Accelerators, Spectrometers, Detectors and Associated Equipment 673 (2012) 107--110.
\newblock \href {https://doi.org/https://doi.org/10.1016/j.nima.2012.01.024} {\path{doi:https://doi.org/10.1016/j.nima.2012.01.024}}.
\newline\urlprefix\url{https://www.sciencedirect.com/science/article/pii/S0168900212000642}

\bibitem{article}
H.~Jansen, S.~Spannagel, A.~Bulgheroni, G.~Claus, E.~Corrin, D.~Cussans, J.~Dreyling-Eschweiler, D.~Eckstein, T.~Eichhorn, M.~Goffe, I.~Gregor, D.~Haas, C.~Muhl, H.~Perrey, R.~Peschke, P.~Roloff, I.~Rubinskiy, M.~Winter, Performance of the eudet-type beam telescopes, EPJ Techniques and Instrumentation 3 (03 2016).
\newblock \href {https://doi.org/10.1140/epjti/s40485-016-0033-2} {\path{doi:10.1140/epjti/s40485-016-0033-2}}.

\bibitem{HIGHLAND1975497}
V.~L. Highland, \href{https://www.sciencedirect.com/science/article/pii/0029554X75907430}{Some practical remarks on multiple scattering}, Nuclear Instruments and Methods 129~(2) (1975) 497--499.
\newblock \href {https://doi.org/https://doi.org/10.1016/0029-554X(75)90743-0} {\path{doi:https://doi.org/10.1016/0029-554X(75)90743-0}}.
\newline\urlprefix\url{https://www.sciencedirect.com/science/article/pii/0029554X75907430}

\end{thebibliography}

\end{document}